\documentclass[11pt,twoside]{article}
\usepackage{asp2010}

\resetcounters

\begin{document}

\title{PHL~1811 Analogs: A Population of X-ray Weak Quasars}
\author{Jianfeng~Wu,$^1$ W.~N.~Brandt,$^1$ P.~B.~Hall,$^2$ R.~R.~Gibson,$^3$ G.~T.~Richards,$^4$ 
D.~P.~Schneider,$^1$ O.~Shemmer,$^5$ D.~W.~Just,$^6$ and S.~J.~Schmidt$^3$
\affil{$^1$The Pennsylvania State University, University Park, PA 16802, USA}
\affil{$^2$York University, Toronto, ON M3J 1P3, Canada}
\affil{$^3$University of Washington, Seattle, WA 98915, USA}
\affil{$^4$Drexel University, Philadelphia, PA 19104, USA}
\affil{$^5$University of North Texas, Denton, TX 76203, USA}
\affil{$^6$Steward Observatory, Tucson, AZ 85721, USA}}
%\affil{$^1$Department of Astronomy \& Astrophysics, The Pennsylvania State University, University Park, USA}
%\affil{$^2$Department of physics \& Astronomy, York University, Toronto, Canada}
%\affil{$^3$Department of Astronomy, University of Washington, Seattle, USA}
%\affil{$^4$Department of Physics, Drexel University, Philadelphia, USA}
%\affil{$^5$Department of Physics, University of North Texas, Denton, USA}
%\affil{$^6$Steward Observatory, University of Arizona, Tucson, USA}}
%\affil{$^1$Department of Astronomy \& Astrophysics, The Pennsylvania State University, University Park, PA 16802 USA}
%\affil{$^2$Dept. of physics \& Astronomy, York University, Toronto, ON M3J 1P3, Canada}
%\affil{$^3$Department of Astronomy, University of Washington, Seattle, WA 98915, USA}
%\affil{$^4$Department of Physics, Drexel University, Philadelphia, PA 19104, USA}
%\affil{$^5$Department of Physics, University of North Texas, Denton, TX 76203, USA}
%\affil{$^6$Steward Observatory, University of Arizona, Tucson, AZ 85721, USA}}

\begin{abstract}
We report on a population of \hbox{X-ray} weak quasars with similar UV emission-line properties 
to those of the remarkable quasar PHL~1811. All \hbox{radio-quiet} PHL~1811 analogs are notably 
\hbox{X-ray} weak by a mean factor of $\approx13$, 
with hints of heavy \hbox{X-ray} absorption. Correlations between the \hbox{X-ray} weakness and UV 
emission-line properties suggest that PHL~1811 analogs may have extreme wind-dominated broad emission-line 
regions (BELRs). We propose an AGN geometry that can potentially unify the PHL~1811 analogs and the 
general population of weak-line quasars. 
\end{abstract}

Luminous \hbox{X-ray} emission is a nearly universal property of efficiently accreting supermassive black holes. 
However, there are a few examples of active galactic nuclei (AGNs) 
that emit \hbox{X-rays} much more weakly. An extreme case is the optically bright quasar PHL~1811 ($z=0.192$). 
It is consistently \hbox{X-ray} weak by a factor of \hbox{$\approx30$--$100$} compared to typical quasars 
with similar UV luminosity (Leighly et~al. 2007a). This quasar also has unusual UV emission-line properties: 
weak and strongly blueshifted high-ionization lines 
(e.g., C~{\sc iv}), weak semi-forbidden lines (e.g., C~{\sc iii}]) and forbidden lines (e.g., [O~{\sc iii}]), 
and strong UV Fe~{\sc ii/iii} emission (Leighly et~al. 2007b).

\begin{figure}[!t]
\plotone{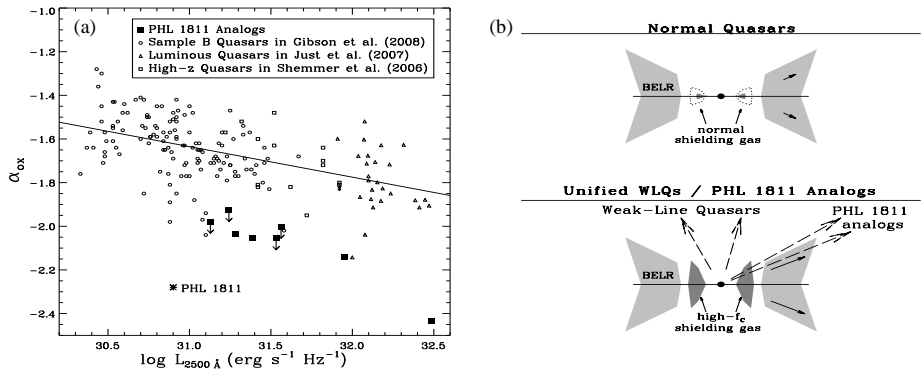}
\caption{\footnotesize{{\bf (a):} $\alpha_{\rm ox}$--$L_{2500~\mathring{\rm{A}}}$ diagram for \hbox{radio-quiet} PHL 1811 
analogs (filled squares) and a comparison sample of typical \hbox{radio-quiet} quasars without broad absorption lines 
from the literature (open symbols). Downward arrows represent $\alpha_{\rm ox}$ upper limits. The solid line shows 
the $\alpha_{\rm ox}$--$L_{2500~\mathring{\rm{A}}}$ correlation in Just et~al. (2007). 
{\bf (b):} Schematic illustration of the WLQ/PHL~1811 analog unification hypothesis.
Each panel is a side view of an accretion disk around a black hole. The \hbox{X-ray}
continuum source is close to the black hole. Normally, the shielding gas
only covers part of the BELR (top panel). When the shielding gas has a high BELR covering factor (bottom panel), 
it absorbs most of the \hbox{X-ray} photons, and a strong wind
is generated in the BELR. When such a quasar
is viewed through the shielding gas, an \hbox{X-ray} weak PHL~1811 analog is seen.
When such a quasar is viewed away from the wind direction,
an \hbox{X-ray} normal WLQ is seen.}}
\end{figure}

To determine if there is a {\it population} of such \hbox{X-ray} weak quasars
and to explore the connection between their UV emission-line
properties and \hbox{X-ray} properties, we conducted a {\it Chandra} survey of a 
sample of Sloan Digital Sky Survey (York et~al. 2000) quasars with similar 
UV emission-line properties to those of PHL~1811 (Wu et~al. 2011a).
Their relative \hbox{X-ray} brightness is quantified by the
$\Delta\alpha_{\rm ox}$ parameter.\footnote{$\alpha_{\rm ox}$ is defined to be the slope of an
assumed power law connecting the rest-frame 2500~\AA\ and 2~keV
luminosities, i.e., $\alpha_{\rm ox}=0.384 \log(L_{\rm
2~keV}/L_{2500~\mathring{\rm{A}}})$. This quantity is well known
to be correlated with $L_{2500~\mathring{\rm{A}}}$ (e.g., Just
et al. 2007; see Fig. 1a). We also define $\Delta\alpha_{\rm ox}$ =
$\alpha_{\rm ox}$(observed)$-$$\alpha_{\rm ox}$($L_{2500~\mathring{\rm{A}}}$), which
quantifies the observed \hbox{X-ray} luminosity relative to that expected
from the $\alpha_{\rm ox}$-$L_{2500~\mathring{\rm{A}}}$ relation.} All
\hbox{radio-quiet} PHL~1811 analogs, without exception, are remarkably
\hbox{X-ray} weak by a mean factor of $\approx13$ 
(mean $\Delta\alpha_{\rm ox}=-0.42\pm0.05$; see Fig.~1a), which shows the 
connection between soft spectral energy distributions (\hbox{X-ray} weak and optical/UV 
strong) and PHL~1811-like UV emission lines. Their $\Delta\alpha_{\rm ox}$ values 
are correlated with the C~{\sc iv} equivalent width and blueshift, 
showing our sample may represent quasars whose
BELRs are dominated by AGN winds (see Richards et~al.
2011). Our \hbox{radio-quiet} PHL~1811 analogs show a hard average \hbox{X-ray}
spectrum ($\Gamma=1.1^{+0.45}_{-0.40}$) which may indicate heavy \hbox{X-ray}
absorption. Observationally, the PHL~1811 analogs appear to be a subset 
($\approx30\%$) of weak-line quasars (WLQs; e.g., Diamond-Stanic 
et~al. 2009; Shemmer et~al. 2009). We propose an AGN geometry that can
potentially unify the PHL~1811 analogs and the general WLQ
population via orientation effects (see Fig.~1b). This model
assumes a subset of quasars in which the highly ionized
``shielding gas'' (e.g., Murray et~al. 1995) covers most of the BELR
and blocks the \hbox{X-ray} photons from reaching the BELR, resulting in
weak high-ionization lines and strong wind acceleration. Our recent work on 
low-redshift WLQs (Wu et~al. 2011b) provides support to this unification model.
%(and thus highly blueshifted high-ionization lines). 
%This model assumes a subset of quasars in which the highly ionized ``shielding gas''
%(see Murray et~al. 1995) covers most of the BELR and blocks the \hbox{X-ray} photons,
%resulting in weak high-ionization lines and strong wind acceleration. 

%\acknowledgements We gratefully acknowledge financial support from {\it Chandra} 
%X-ray Center grant GO0-11010X (J.W., W.N.B.), NASA ADP grant NNX10AC99G (J.W., W.N.B.), and 
%NSERC (P.B.H.). The SDSS website is http://www.sdss.org/.

\end{document}